# Shear Wave Filtering in Naturally-Occurring Bouligand Structures

Nicolás Guarín-Zapata[1], Juan Gomez[2], Nick Yaraghi[3], David Kisailus[3, 4], Pablo D. Zavattieri[1,*]

[1]*Lyles School of Civil Engineering, Purdue University, West Lafayette, IN 47907, USA*
[2]*Civil Engineering Department, Universidad EAFIT, Medellín, 050022, Colombia*
[3]*Materials Science and Engineering, University of California, Riverside, Riverside, CA 92521, USA.*
[4]*Department of Chemical and Environmental Engineering, University of California, Riverside, Riverside, CA 92521, USA*

**Abstract:** Wave propagation was investigated in the Bouligand-like structure from within the dactyl club of the Stomatopod, a crustacean that is known to smash their heavily shelled preys with high accelerations. We incorporate the layered nature in a unitary material cell through the propagator matrix formalism while the periodic nature of the material is considered via Bloch boundary conditions as applied in the theory of solid state physics. Our results show that these materials exhibit bandgaps at frequencies related to the stress pulse generated by the impact of the dactyl club to its prey, and therefore exhibiting wave filtering in addition to the already known mechanisms of macroscopic isotropic behavior and toughness.

## 1. Introduction

Many biological organisms are known for their ability to produce hierarchically arranged materials from simple components, resulting in structures that provide mechanical support, protection and mobility. These structures are used to perform a wide variety of functions ranging from structural support and protection to mobility and other basic life functions. All of this is done using only the minimum quantities of a limited selection of constituent materials [1–3], synthesized under mild conditions. The diversity and multifunctionality identified in these materials, combined with their robust mechanical properties [2] make them a rich source of inspiration for the design of new materials. One particular example of a natural material with impressive mechanical properties can be found in the dactyl club of stomatopods [1], [4–7]. The *stomatopods* (or mantis shrimps), are an ancient group of marine tropical and subtropical crustaceans that are, on average, 15 cm long but can reach lengths of nearly 40 cm. A distinct feature of stomatopods versus other crustaceans is the presence of a pair of thoracic appendages that are specifically adapted for close-range combat (See Fig. 1a, b). Stomatopods are divided into two groups, depending on the shape of these appendages: those that hunt by impaling their prey with spear-like structures (*spearers*), and those that smash them with a powerful blow from a heavily mineralized club (*smashers*) [5], [8–10]. The dactyls of spearers contain spiny appendages with barbed tips that prevent prey from slipping off. On the other hand, dactyls from smashers have a hammer-like structure[11] that can reach, upon impact, accelerations as high as 10400*g* and speeds close to 23 m/s, generating forces up to 1500 N [10]. This hammer-like composite structure can inflict considerable damage after impact against the wide variety of heavily mineralized biological structures present in its preys. In fact, this is reflected in its diet. Smashers, feed on armored animals such as snails, hermit crabs, clams and crabs, which they batter to pieces [11]. Despite these significant forces, the dactyl clubs are fracture-resistant and are able to tolerate thousands of such blows. This astounding capacity to tolerate stress waves generated from the impact of the dactyl club against its preys has prompted questions about the underlying mechanisms responsible for such a high strength to sustain dynamic loads [1], [7].

---

[*] Corresponding author: zavattie@purdue.edu

Previous research, conducted by Weaver et al. [7] aimed at identifying the microstructural features of the dactyl club in stomatopods, have recognized its structure as a multiregional biological arrangement made of an external layer, called the impact region, supported by a striated and a periodic zones. These are described in Figure 1b-1e where we show a description of a transversal cross section of the dactyl club together with its microstructural features shown at increasing levels of resolution. In particular Fig. 1c shows a schematic representation of a cross section of the dactyl club where we have labeled the impact, periodic and striated regions as (I), (II) and (III) respectively. The impact and periodic regions are also observed in the optical micrograph of a polish cross section of the dactyl club (Fig. 1d). The layered nature is more evident in the periodic region. A closer examination reveals that this periodicity is related to its helicoidal arrangement (also known as Bouligand structure) of unidirectional chitin fibrils surrounded by amorphous mineral [7]. This structure is observed in the scanning electron microscope (SEM) image shown in Fig. 1e and in the idealized model (see Fig. 1f) where we introduce fibers mimicking such a helicoidal arrangement. It should be noted that each line in Fig. 1d correspond to a complete 180º rotation of the fibers in Fig. 1e. While it is not clear from Fig. 1d, the impact region also exhibits a similar arrangement of fibers.

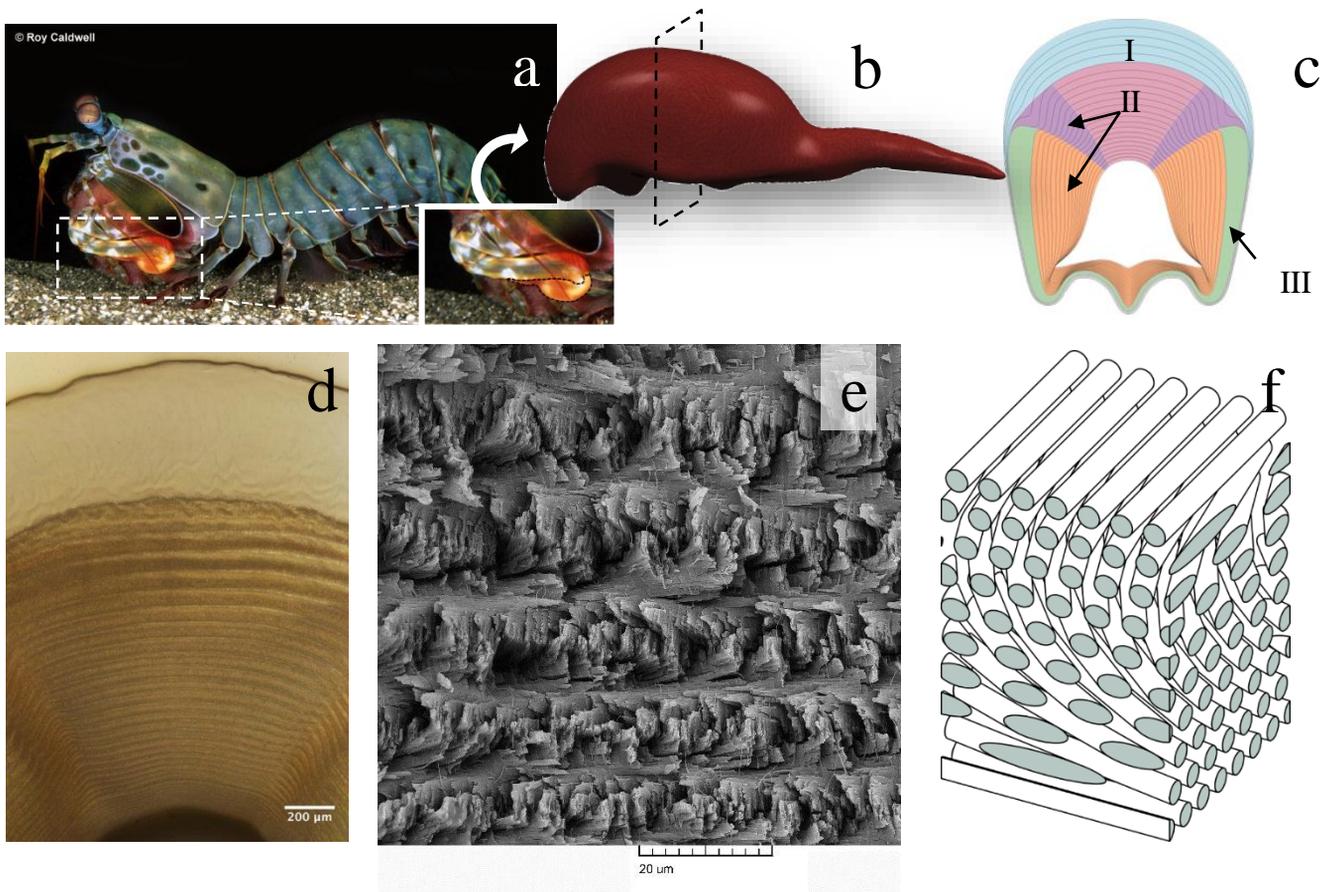

**Figure 1:** Hierarchical structure of the dactyl club of the stomatopod. a) Image of the smashing peacock mantis shrimp (Odontodactylus Scyllarus). b) Model of the dactyl club c) Schematic of a transverse section of the dactyl club highlighting the (I) Impact region (light blue) (II) Periodic region (pink, purple, orange) and (III) Striated region (green). d) Optical micrograph of a polished transverse section of the impact and periodic region, e) SEM micrograph of a fractured surface of the periodic region highlighting the helicoidal microstructure. f) Model of the Bouligand microstructure.

The nearly periodic nature of the microstructure in the dactyl club seems to suggest that the interaction between the microstructure and propagating stress waves can lead to phenomena that are common in phononic crystals and metamaterials, such as bandgaps and dispersion of waves [12–14]. As such, manipulating the microscopic structure and geometry of a crystal-like solid, the macroscopic behavior is changed and a media capable of molding and manipulating waves can be conceived [12–16]. The field of phononic-crystals has some aspects in common with biological materials; e.g., the overall behavior is determined by the presence of microstructural features. Furthermore, the approach is not just one of changing the properties of the different phases but with the architecture (topology/geometry). As such, the aim of this paper is to explore the microstructural properties of the periodic and impact region of the dactyl club in stomatopods from the perspective of spatial periodicity and wave propagation properties in order to identify whether bandgaps and dispersion of waves could be one of the contributing mechanisms responsible for its remarkable impact tolerance capability.

Phononic crystals are materials with a periodic repetition of a unit cell, which results in the periodicity of its mechanical properties, i.e., elastic moduli and density. The wave propagation in this case corresponds to an elastic disturbance [16], [17]. One of the main properties of a phononic crystal is the possibility of exhibiting band gaps, i.e., frequency ranges where waves are forbidden to propagate [16], [18]. The name of *phononic crystals* has been coined from the field of photonic crystals in optics, and both, phononic and photonic crystals can be studied using concepts extracted from the theory of solid-state physics. In particular, the use of Bloch's theorem [18], [19] allows us to determine the material band gaps after studying a single unitary material cell. The differences between electronic/photonic/phononic crystal reside in the equations that are being solved: Schrödinger equation, Maxwell equation or Navier-Cauchy equations, for *electronic*, *photonic* and *phononic* crystals, respectively [16], [19], [20]. In all the cases, the media have properties exhibiting space periodicity. There is a wide variety of applications in the field of elastodynamics [16]. Hladky-Hennion and Decarpigny presented a study applying a finite element method (FEM) to periodic materials used as coatings to avoid detection of submarines by ultrasound waves [21]. Ruzzene et al. modeled honeycombs and re-entrant honeycombs (hexagons with inverted angles) to find the directionality of the material (that can be used as an acoustic filter) [22]. Wang et al. developed a material with tunable band gaps, controlling local instabilities in the microstructure [15].

The existence of helically stacked plies (or Bouligand structures) have already been noticed and investigated in terms of their microstructural features in a variety of other animals, such as fish scales [23], exoskeletons of beetles [24], crabs [25] and lobsters [26–29]. For instance, Sachs and co-workers presented experimental measurements for lobster's cuticle using digital image correlation, obtaining the behavior for the elastic and plastic regimes [27–29]. Nikolov and co-workers have studied the hierarchical composition of the cuticle in the exoskeleton of the American lobster [30–32] and showed that the level of anisotropy in the elastic properties of the cuticle is very high at the nano-scale. However, such anisotropy decreases monotonically moving to the higher scale and exhibiting almost isotropic elastic properties at the millimeter scale. A similar result was reported in the exocuticles of beetles by Sun and Bhushan, who characterized the structure and mechanical properties of beetle wings and reported their lightweight nature; high strength; superhydrophobicity and structural coloration [33]. Zimmerman and co-workers showed that a Bouligand-type arrangement existing in Arapaima gigas fish scales can adapt to loads in different orientations by reorienting in response to different stresses [23]. On the other hand, Bouligand structures have also been studied from a wave propagation point of view. For instance Vukusic and Sambles presented a natural photonic structure in hawkmoths exhibiting low-reflectance suitable for stealth technology [34]. Campos-Fernández measured the reflection spectra of beetles and proposed a multilayered model to explain the variations in the obtained spectra [35], while Zhang and To used a biomimetic multilayered model with a hierarchical layered structure to obtain broadband wave filtering in phononic crystals [36].

In this work we adopt a layered model for the unit cell in the periodic and impact regions of the dactyl club. We first discuss the different length scales present in a typical impact pulse and, particularly, its relation with the microstructural features encountered in the dactyl club to justify our choice of a model based on a unitary material cell with rotated layers of a transverse isotropic material. For completeness we also describe briefly the wave equations for unbounded media leading to the form of the Christoffel wave equations [37–39] after considering propagating plane waves. The proposed model for the unitary cell is then analyzed with the combination of a propagator matrix approach, introduced by Yang et al [40], to study layered materials, together with the imposition of Floquet-Bloch periodic boundary conditions [18], [19]. From these combined analyses we determine dispersion relations, and particularly, frequency bandgaps. Subsequently, different assumed geometrical parameters of the unit cell which are compared with a stress pulse representative of a typical impact sustained by the stomatopod. As a performance measure of the different microstructural configurations we introduce a scalar parameter quantifying the amount of transmitted energy through the different bandgaps.

## 2. Modeling

We now describe relevant assumptions and modeling aspects used in the study of the dynamical response of the material present in the dactyl club in the stomatopod. We will focus in the response of a unitary material cell conformed by an arrangement of stacked layers of uniaxial fibers (see Figure 1f). Although such arrangements are quite common in composite materials [41], helicoidal configurations of uniaxial fibrils are unique of biological materials. One fundamental aspect in our analysis technique is the fact that when the ratio between the characteristic wavelength $\lambda$ of the input excitation to the intrinsic length scale $\ell$ representative of the relevant microstructural features in the dactyl club (i.e., fiber diameter and intra-fiber distance) is large, each layer of uniaxial fibrils can be represented as a homogeneous transverse isotropic material. That assumption has been previously used by Yang et al. to study the propagation of mechanical waves in an arrangement of uniaxial plates via a propagator matrix approach [40] and by Varadan et al. who used nondestructive ultrasonic testing measurements to determine the direction of the major and minor axes of polarization in a helicoidal composite and find a match between computed and experimental results [42]. This assumption has been used in previous theoretical and experimental biomimetic studies of arthropods [24], [43], [44]. Furthermore, for the case of the American Lobster, Nikolov and workers [31], [32] presented that the anisotropy in the elastic properties of the cuticle decreases monotonically from small to large length scales. In fact, for the scale of interest in this paper (i.e., layers for the Bouligand stacking), their results indicated that the material is mostly transversely isotropic.

### 2.1 Length scale considerations

To properly analyze the characteristic length scales of the problem, i.e., the wavelength and internal microstructural characteristic dimension, we take information from the dynamic impact event reported in previous works [5], [7], [10] and compare it to the geometric features identified in the periodic region. The optical micrograph in Figure 2a depicts a transverse cross section of the periodic and impact regions of the dactyl club. Each line correspond to a complete a 180º-rotation of the fibers [1], [7]. The distance between lines, also called pitch distance, $D$, is plotted in Fig. 2b along the *y*-direction (where $y = 0$ is the interface between the impact and periodic region) for three different specimens. The impact region was included for one specimen as a reference. As it can be observed, both impact and periodic region display a pitch distance gradient varying from 140 μm in the impact region to less than 10 μm in the bottom of the periodic region. In some cases, a local maximum in the periodic region near the interface with the impact region is observed. However, even if the values of $D$ have some slight variation with specimen, they show the same trend.

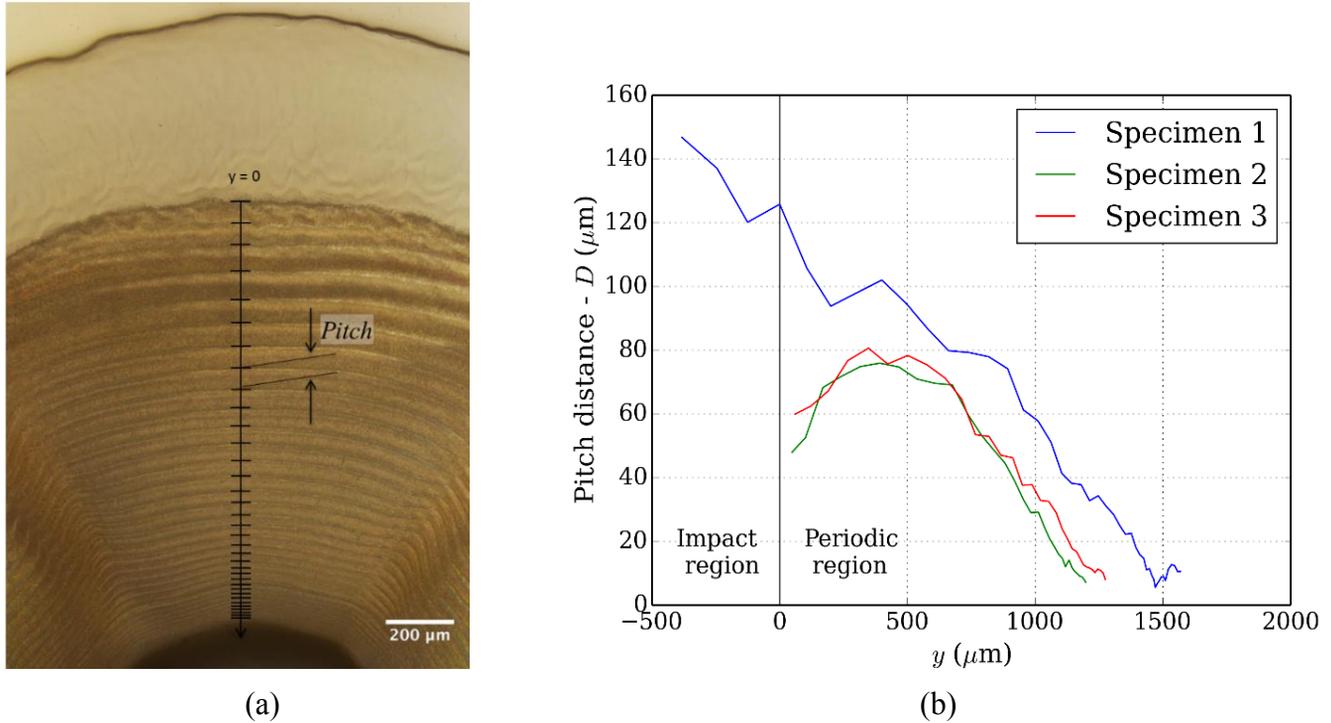

(a)          (b)

**Figure 2:** Pitch gradient in the periodic region. a) Optical micrograph. b) Pitch distance in the periodic region.

It has been shown by previous experimental [10] and numerical investigations [7] that the impact, represented in terms of a stress pulse, has a duration that ranges from $\Delta t \approx 1.4 \times 10^{-2}$ ms to $\Delta t \approx 1.6 \times 10^{-1}$ ms. In this work, we employ the shortest duration, which translates in a signal rich in high frequency content. In order to represent an extreme scenario, we also assume a square shaped pulse based on the fact that the use of a time discontinuity in stress requires more high frequency terms than the smooth curves measured in [7], [10]. Figures 3a and 3b describe time and frequency domain comparisons between our theoretical square pulse and the impact measurement reported in [7]. It is observed how the intensity of the Fourier spectrum decreases with frequency in such a way that if we take $\omega \Delta t = 100$, we are already considering approximately 99% of energy in the pulse. Using this approach, we can estimate a cutoff frequency of $11.4 \times 10^{14}$ Hz. For wave speeds in the range of 500-5000 m/s, this frequency yields a characteristic wavelength $\lambda$ that will range between 438 and 4380 μm. Considering an average pitch distance of 75 μm near the impact region, the ratio $D / \lambda$ ranges from 0.017 to 0.17. Finally, high resolution SEM studies reported by Grunenfelder et al. [1] revealed the presence of ~ 1.13 μm mineralized fiber bundles consisting of individual fibrils ranging from 20 to 50 nm in diameter. Thus, we assume that the ratio between $\lambda$ and the fibers characteristic length $\ell$ (i.e., fiber diameter and inter-fiber distance) is sufficiently large to consider each layer of unidirectional fibers as a homogeneous transverse isotropic material.

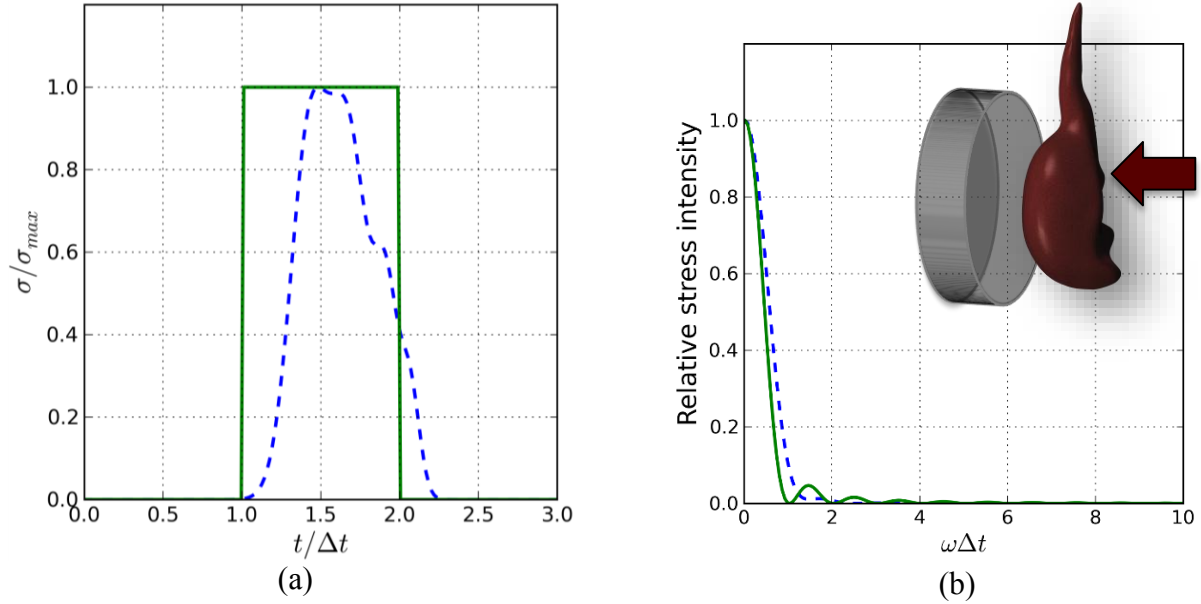

(a) (b)

**Figure 3:** a) Impact pulse from the stomatopod's dactyl club strike on a flat target b) power spectra. The dashed line represents the waveform for the pulse simulated Weaver et al. [2012]. The solid line represents a square pulse that represents higher high-frequency content. Inset: schematics of the dactyl club impacting a target as performed in previous experimental and modeling works [7], [10].

## 2.2 Wave equation for elastic materials

The particle motion in an elastic solid is governed by the following set of partial differential equations:

$$\sigma_{ij,i} + b_i = \rho \ddot{u}_j, \qquad (1)$$

where $u_j$, $\sigma_{ij}$ and $b_i$ are the components of the displacement vector, the stress tensor and the body forces vector, while $\rho$ represents the material mass density. In Equation (1) and throughout the paper repeated indexes are implied to obey the summation convention unless specifically stated otherwise. At the same time a comma refers to spatial derivatives while dotted variables describe time derivatives. As discussed in [7], a single impact event is unlikely to induce significant damage or large deformation. Any accumulate damage in the dactyl club is the result of multiple impact through a long period of time. Since our interest is to study a single impact event, the material is considered to satisfy a stress-strain relationship in the form of Hooke's law as:

$$\sigma_{ij} = c_{ijpq} u_{i,j}, \qquad (2)$$

where $c_{ijpq}$ is the stiffness tensor comprising 5 different parameters in the case of a transverse isotropic material [45]. In our particular case, the values of $c_{ijpq}$ are based on [7], [27], [31] for the periodic region of the dactyl club. In a problem involving unbounded media where interest lies in free wave motion, body forces are neglected and a solution with the following space-time representation is assumed:

$$u_j = U_j e^{ik(n_r x_r - v_p t)}. \qquad (3)$$

In (3) $U_j$ is the amplitude of the displacement vector in the $j$ direction, $k$ is the wave number, $n_r$ is the propagation direction, and $v_p$ is the phase speed of the wave. We assume a time dependence of the form $e^{-i\omega t}$, which is omitted hereafter and where $i$ is the imaginary unit and $\omega$ is the circular frequency. Substitution of (2) and (3) into (1) gives the Christoffel wave equation [37–39]:

$$[\Gamma_{ik} - \rho v_p^2 \delta_{ik}]U_k = 0, \tag{4}$$

where $\Gamma_{ik} \equiv c_{ijpq} n_j n_q$ is the so-called Christoffel stiffness tensor, while $\delta_{ik}$ is the identity tensor. Equation (4) has no-trivial solution if:

$$\det[\Gamma_{ik} - \rho v_p^2 \delta_{ik}] = 0, \tag{5}$$

which leads to the existence of three types of bulk waves [37], [39] namely: a quasi-P wave (with polarization direction almost equal to the propagation direction), a quasi-S wave (with polarization direction almost perpendicular to the propagation direction), and an S-wave (polarized orthogonal to the quasi-S wave and to the direction of propagation). The solution of Equation (5) gives the phase speed for each type of polarization, and, consequently, the speed of each wave depends on the direction of propagation.

**2.3 Layered Model**

A material element cell is represented by an arrangement of $N$ stacked layers of unidirectional fibers embedded in a matrix of thickness $d$ and a pitch distance $D$. Such material element cell represents the behavior of the bulk Bouligand structure inside the dactyl club. As such, our main assumptions are: (1) the radius of curvature of the layers is much larger than $D$, and (2) there are enough layers to ignore boundary effects from the impact region or the inner portion of the dactyl club. Figure 4 shows the material element cell in the form of $N$ helicoidally stacked layers of fibers representative of the periodic region of the dactyl club already shown in Figure 1f. It should be noted that this model does not explicitly consider one fiber across the thickness of an individual layer as depicted in Fig. 4. Instead, the presence of randomly distributed, but unidirectional chitin fibrils surrounded by amorphous mineral are accounted in the elastic behavior of a homogeneous transversely isotropic layer. Each $n^{th}$ layer forms an angle $\alpha_n = (n-1)\alpha$ with the global $x$-axis, where $n$ is the layer number and $\alpha$ is the pitch angle formed between two adjacent layers of unidirectional fibers. The $N$ layers complete a 180° rotation through a pitch distance $D=Nd$.

Helicoidal materials as the one described in Figure 4 are known to exhibit shear wave filtering [40], [42], [46] thus we use such a model to quantify these filtering capabilities in the case of the dactyl club of stomatopods. We analyze the response of the material cell in terms of dispersion relations for a range of values of the mechanical material parameters and microstructural dimensions. The resulting curves are then compared with the Fourier spectra for the input pulse. Since the material has been modeled as stacked (homogeneous) transverse isotropic layers, we adopt as solution method the propagator matrix [47] where the three-dimensional partial differential equations are converted into a system of ordinary differential equations using constitutive equations and continuity between layers. Moreover, the periodic nature of the unit cell is considered through the imposition of Bloch periodic boundary conditions. It has been suggested that between 10 and 50 unit cells the periodicity of the material can be taken into account [17], in our case we have between 30 and 50.

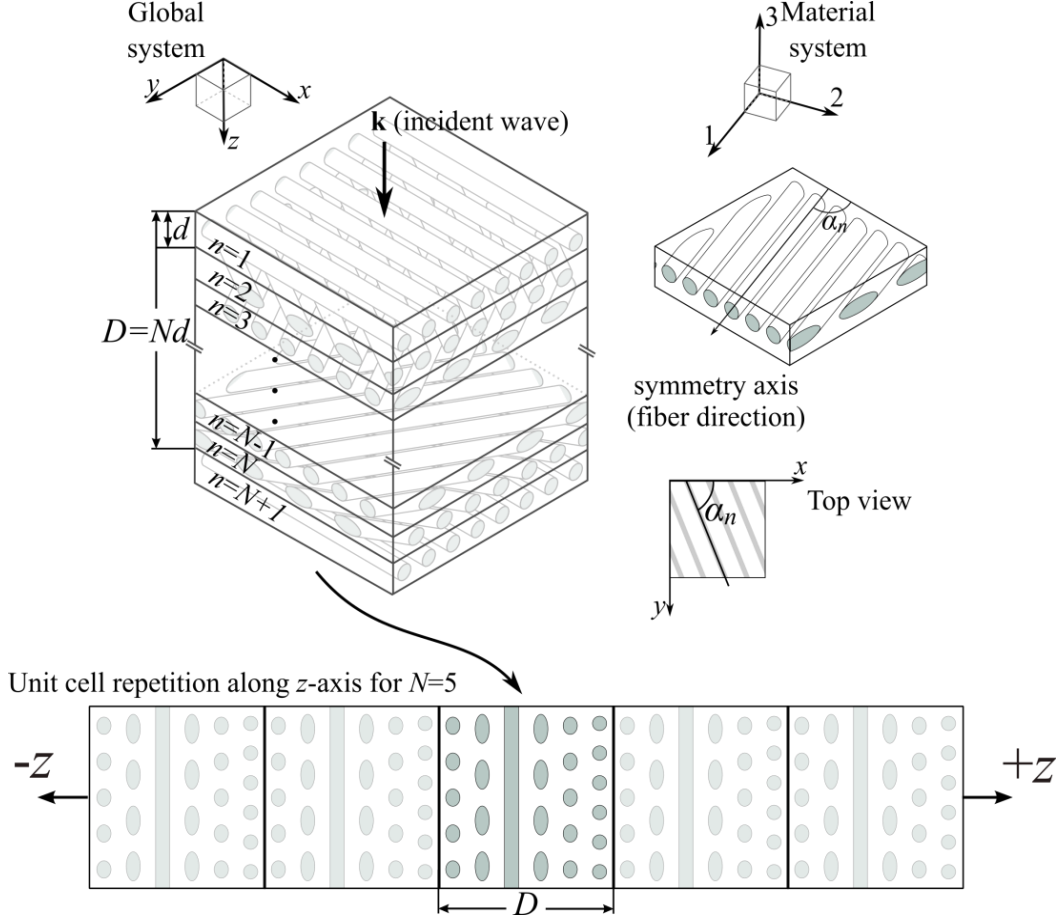

**Figure 4:** Schematic of the layered model. The parameter, $D$ is the pitch distance (i.e., the distance to make a rotation of 180º); $d$ is the layer thickness; N is the number of layers in each pitch; and $\alpha_n$ is the angle for the nth layer, i.e., $\alpha_n = (n\text{-}1)\,\alpha$. (right) Orientation of a layer with respect to the global coordinate system. The material coordinate system is depicted in the upper right corner, 1 is the direction along the fiber primary axis while 2 and 3 are both perpendicular to 1. In the bottom we show how each unit cell is repeated along the $z$-axis to form the periodic material as a whole.

The equation of conservation of linear momentum without body forces admits a solution of the form

$$u_r(x,z) = U_r(z)\exp(-ik_x x) \qquad (6)$$

$$\sigma_{pq}(x,z) = S_{pq}(z)\exp(-ik_x x), \qquad (7)$$

where $U_r$ is the amplitude of the displacement vector in the $r$ direction, $S_{pq}$ is the amplitude of the $pq$ component of the stress tensor and where $k_x$ is the $x$ component of the wave vector. Unless explicitly stated otherwise all the indexes refer to components in the global coordinate system. Substitution of (6) and (7) into (1) yields after conducting some algebraic manipulations the equation of motion for each layer in the form of a system of ordinary differential equations as [40]:

$$\frac{dV(z)}{dz} = i[P]V(z), \qquad (8)$$

and where $V(z) = [U_x, U_y, U_z, S_{xz}, S_{yz}, S_{zz}]^T$ is a state vector given by the 3 components of the displacement vector and the 3 components of the stress tensor that need to be continuous across the interfaces. In this equation [P] is a coupling 6×6 matrix that contains the material information of the composite and which explicitly reads:

$$[P] = \begin{bmatrix} 0 & 0 & k_x & \frac{-iC_{44}}{\Delta} & \frac{-iC_{45}}{\Delta} & 0 \\ 0 & 0 & 0 & \frac{-iC_{45}}{\Delta} & \frac{-iC_{55}}{\Delta} & 0 \\ \frac{k_x C_{13}}{C_{33}} & \frac{k_x C_{36}}{C_{33}} & 0 & 0 & 0 & \frac{-i}{C_{33}} \\ A & B & 0 & 0 & 0 & \frac{k_x C_{13}}{C_{33}} \\ B & M & 0 & 0 & 0 & \frac{k_x C_{36}}{C_{33}} \\ 0 & 0 & i\rho\omega^2 & k_x & 0 & 0 \end{bmatrix}, \quad (9)$$

with

$$A = i\left[\rho\omega^2 - \left(C_{11} - \frac{C_{13}C_{13}}{C_{33}}\right)k_x^2\right] \quad (10)$$

$$M = i\left[\rho\omega^2 - \left(C_{66} - \frac{C_{36}C_{36}}{C_{33}}\right)k_x^2\right] \quad (11)$$

$$B = -i\left(C_{16} - \frac{C_{13}C_{36}}{C_{33}}\right)k_x^2 \quad (12)$$

$$\Delta = C_{44}C_{55} - C_{45}^2. \quad (13)$$

and where the terms $C_{IJ}$ refer to components of the stiffness tensor ($c_{ijkl}$) represented in Voigt notation, where second and fourth order tensors are collapsed into $R^6$ vectors and 6×6 symmetric matrices respectively [48]. Here the components of the stiffness tensor $C_{IJ}$ are given in the local coordinate system (see Figure 4).

Equation (8) has a general solution of the form:

$$V(D) = [Q]V(0). \quad (14)$$

where the matrix [Q] is defined by

$$[Q] = \exp(i[P_N]d_N)\exp(i[P_{N-1}]d_{N-1})\cdots\exp(i[P_1]d_1), \quad (15)$$

which results after considering displacement and traction compatibility along the interfaces of the N layers in the composite according to the propagator matrix formalism [49], [50]. In (15) $[P_s]$ is the [P] matrix for the $s^{th}$ layer of corresponding thickness $d_s$, and exp refers to the matrix exponential. It is now evident that the matrix [Q] has the information of the material, lamina distribution and angle change in the composite and it establishes a relationship between the state vector at the beginning of the first layer and the state vector at the end of the last layer.

The periodic nature of the composite material found in the dactyl club is now considered through boundary conditions imposed in the form of Bloch's theorem as applied in the theory of phononic crystals [16], [17]. In our context we assume the spatial periodic repetition of the $N$ layers leading to the following relationship involving a phase shift across the cell:

$$V(D)=V(0)\exp(-ikD), \qquad (16)$$

which leads to the following eigenvalue problem

$$[Q]V(0)= \gamma V(0), \qquad (17)$$

with $\gamma=\exp(-ikD)$.

Thus, the proposed representation for the material cell together with the assumption of periodicity translates into an eigenvalue problem for each frequency, $W$, which allows us to find the dispersion curves providing information about the bulk behavior of waves in the material [18], [51]. The specific nature of the eigenvalue $\gamma$ indicates whether the wave is a propagating wave, in which case it implies a change in phase, or if, by contrast, it is an evanescent or decaying wave, where it indicates a change in amplitude. Whenever $|\gamma|=1$, we observe propagating waves of real wave number $k$; when $|\gamma|<1$, we obtain evanescent waves whose amplitude decays in the direction of incidence and finally, when $|\gamma|>1$, we obtain evanescent waves with amplitude dying out in a direction opposite to the direction of incidence. For propagating waves, each frequency $\omega$ in the eigenvalue problem yields a set of wavenumbers $\pm k_p, \pm k_{s_1}, \pm k_{s_2}$ related to longitudinal waves ($k_p$) and transversal waves ($k_{s_1}$ and $k_{s_2}$). Here, the plus sign refers to a wave in the direction of the incident wave while the minus sign refers to a wave in the opposite direction of the incident one. In the case of evanescent waves, the eigenvalues should be given by complex conjugates and the magnitude $|\gamma|$ could be understood as the factor in which the amplitude changes from one cell to the other [50]. This last case is highly relevant in our study since it demarks the region of existence of a bandgap where the material microstructure exhibits actual filtering capabilities.

## 3 Results and Analysis

The primary purpose behind a dispersion analysis is the determination of the potentially existing propagation modes in a material, through the study of a single elementary cell under the assumption of spatial periodicity. Here, these modes are determined for different combinations of mechanical properties and dimensions of the microstructural features (i.e., pitch distance, pitch angle and layer thickness) in the dactyl club of the stomatopod, with the goal of identifying those combinations for which shear wave filtering is most effective under typically sustained impact loads. A dispersion curve gives the variation of circular frequency against wavenumber (or wavelength) for an elementary cell and it provides an objective description of the propagation properties of the material as the analysis reveals not only the phase and group velocities but also the existence of bandgaps or frequency ranges where waves are forbidden to propagate. As discussed previously, here we use the propagation matrix formalism together with Bloch's theorem to consider the spatial periodicity and the layered nature of the elementary material cell conforming the impact region of stomatopods.

Figure 5 compares the normalized frequency (i.e., $\omega d/c\pi$) versus normalized wave number (i.e., $\kappa D/\pi$) dispersion curves for different configurations of helicoidal composites, after assuming values for the relevant mechanical and geometrical parameters of the periodic cell. In particular we used $E_1 = 30$ GPa, $E_2 = 15$ GPa, $G_{12} = 0.7$ GPa, $v_{12} = v_{23} = 0.25$ and a mass density $\rho = 1400$ kg/m$^3$ which correspond to values close to the ones reported in [7] for the periodic region of the dactyl club. The values for the Young modulus and density are taken from [7]. $v_{12}$ and $v_{23}$ were adopted from [31]. These values are also close to the experimental results presented in [27] (where $v_{12} = v_{23} = 0.28$). Given the uncertainty on these properties, we employ a range of $v_{12}$ and $v_{23}$ from 0.25 to 0.3. Regarding the shear modulus, we obtained the order of magnitude from Nikolov et al., and we varied the value from 0.35 to 0.7 GPa (keeping the same order of magnitude) and recompute the transmitted energy.

The dispersion curves correspond to values of the pitch angle $\alpha = [10°, 20°, 30°, 45°, 60°, 90°]$. We used a reference material velocity of c = 4629 m/s defined according to $c^2 = E_1/\rho$ and a layer thickness to wavelength ratio $d/\lambda = 0.021$. In each curve, the presence of a bandgap is highlighted by a shaded gray area completely delimiting the forbidden frequency range. We also added the Fourier spectrum amplitudes to these plots, to identify the effectiveness of each configuration in filtering dynamic energy. A direct comparison between these spectra and a bandgap will reveal zones where the filtering of energy is more effective.

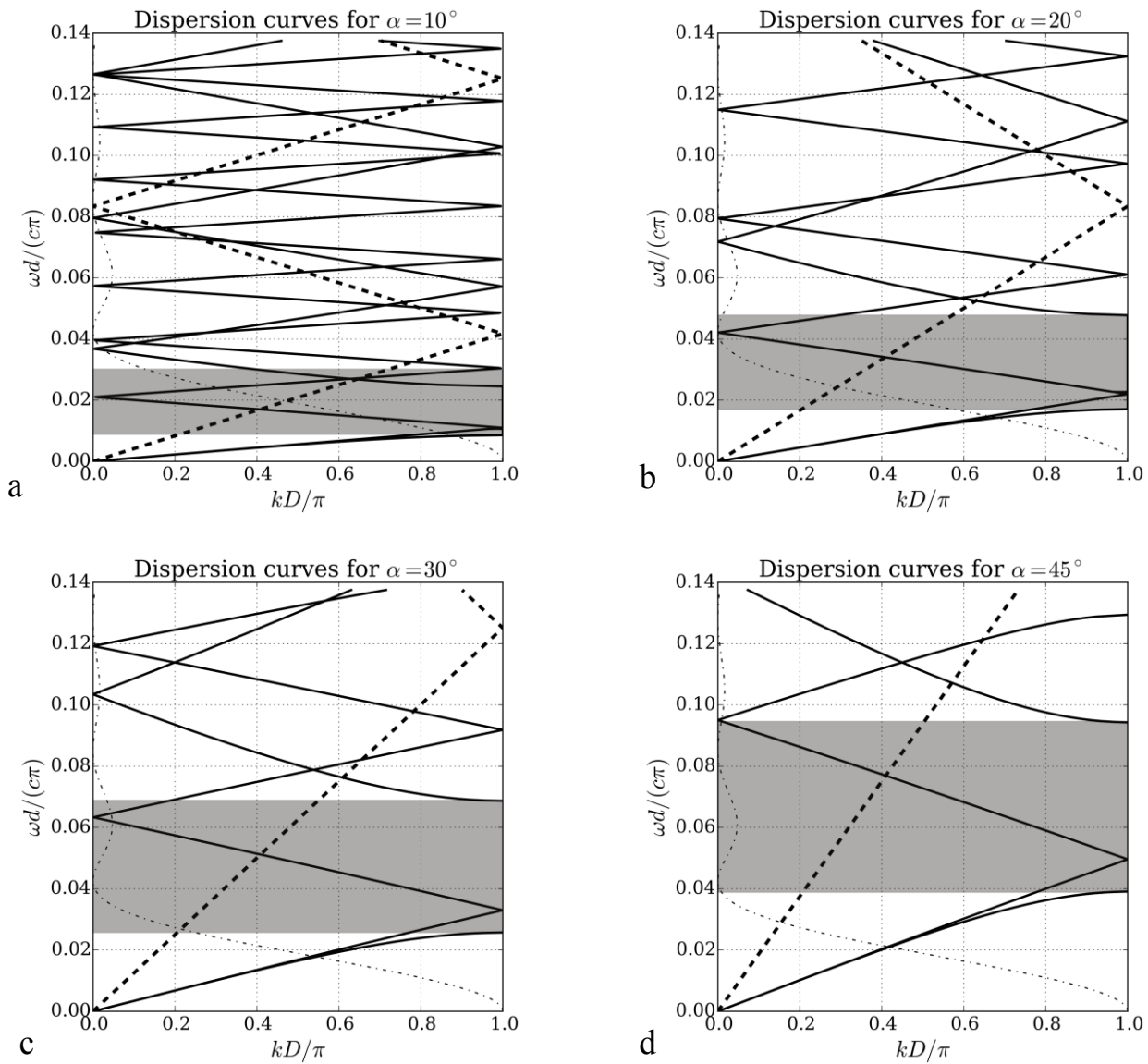

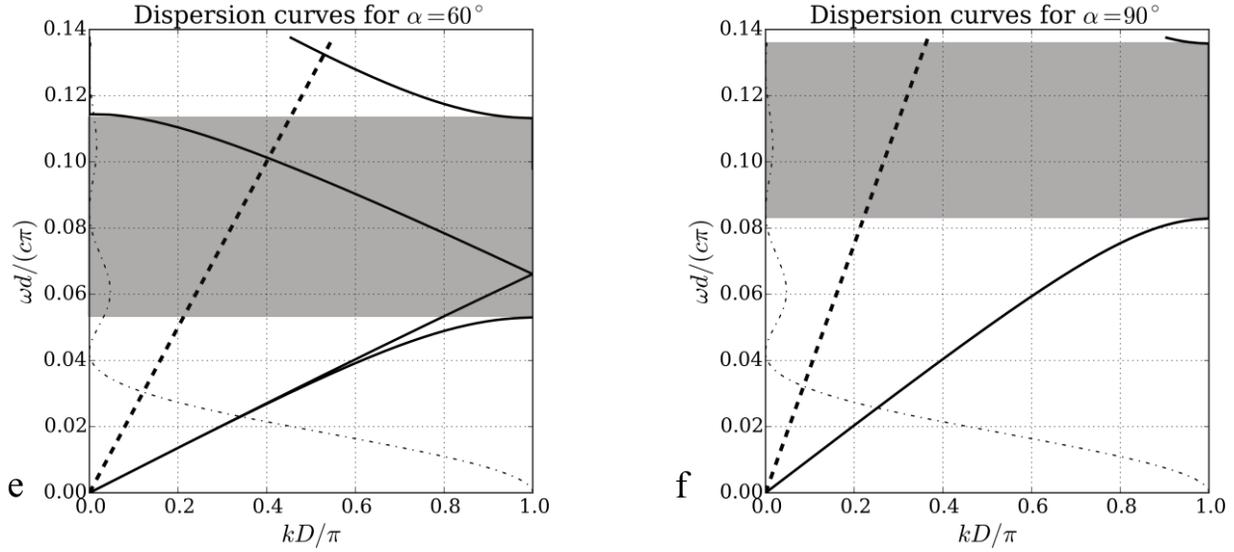

**Figure 5:** Dispersion relations for different parameters values. In all the cases, $d/\lambda=0.021$ and $c=4629$ m/s. The dotted line depicts the spectrum of the impact, the dashed line corresponds to the longitudinal wave, and the solid lines are the quasi-shear waves. We depict the bandgap as the grey region. Angle between layers: a) $\alpha= 10°$, b) $\alpha= 20°$, c) $\alpha= 30°$, d) $\alpha= 45°$, e) $\alpha= 60°$, f) $\alpha= 90°$.

Yang et al. [40], [42] demonstrate that a helicoidal structure is detectable with transverse waves since shear waves would suffer a change in polarization due to rotation in the vibration direction between two adjacent layers (producing a filtering effect). By contrast, the chirality of the composite does not affect the propagation of longitudinal waves. Thus, volumetric incident waves would travel through the material without suffering any dispersion. Since shear and tensile stresses are harmful to the microstructure of the fibrils, filtering these shear waves would have a protective or damage tolerating effect in the inner layers which are highly susceptible to damage. This is helpful to the material since the layers in the dactyl club are generated from the inside out. It is therefore observed from the dispersion curves that the P wave mode (indicated by dashed lines in Fig. 5) is insensitive to the orientation of the fibers due to the perpendicular nature of the incident wave combined to the longitudinal vibration of the particles. By contrast, the transverse, or S wave, modes are dispersive and exhibit bandgaps. The size in the frequency range delimiting the bandgap is also observed to increase as we increase the pitch angle. Such variation in the size of the bandgap is accompanied by a shift in the cutoff frequency towards the high frequency regime. This trade off (between an increase of size and shifting of the bandgap) suggests the presence of a value where the filtered energy is a maximum.

To assess the potential ability of the stomatopod to tolerate the high intensity impact pulse produced upon its daily hunting activity, we now obtain the fraction or amount of energy contained in a single pulse that is actually transmitted by the different microstructural configurations. This is achieved after computing the difference between the total strain energy density imparted by a typical stress pulse and the amount of energy lying inside a band gap bounded by frequencies $\omega_1 - \omega_2$ and associated to a specific microstructural configuration. This transmitted energy, shown as the shaded area in the schematic representation of Fig. 6, is given as

$$E_{trans} = \frac{2}{Y}\left[\int_0^{\omega_1}|\sigma(\omega)|^2 d\omega + \int_{\omega_2}^{\infty}|\sigma(\omega)|^2 d\omega\right] \qquad (18)$$

where *Y* is a conveniently selected elastic moduli of the material which has been introduced for dimensional consistency.

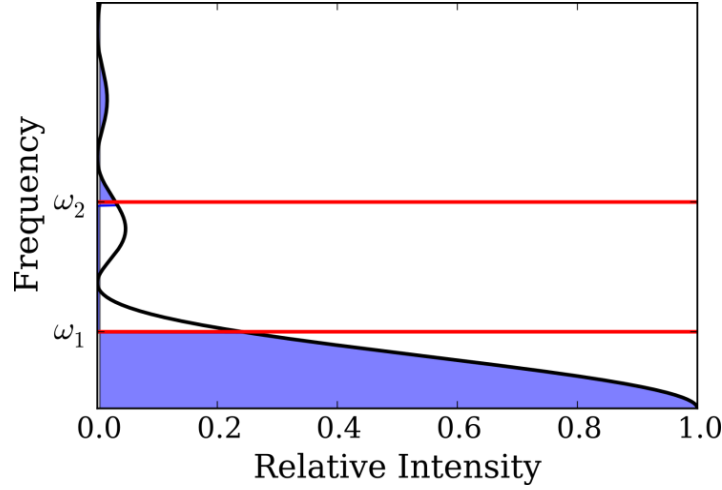

**Figure 6:** Regions for *passing* and *no-passing* energy. The *passing* energy is computed as the shaded region over all the area under the curve. The band gap is denoted by frequencies $\omega_1$ and $\omega_2$ denoted by the horizontal (red) lines.

The performance of the different micro-structural configurations is compared on the basis of the fraction of transmitted energy $\eta$ (or ratio between the transmitted energy and the total energy imparted by a single pulse. $E_{total}$) defined as:

$$\eta = \frac{E_{trans}}{E_{total}} \tag{19}$$

The elastic moduli *Y* in Eq. 18 is the same for $E_{total}$. Figure 7 compares the performance of the resulting microstructures from different perspectives. Here, we introduce two dimensionless frequencies, $\Omega = Df/(2\pi c) = D/\lambda$ and $\theta = d\omega/(2\pi c) = d/\lambda$ which define ratios between the relevant microstructural lengths and the wavelength of the propagating wave. The value used to normalize in Fig. 7 is $\lambda = 438$ μm. These dimensionless frequencies satisfy the relationship $\Omega = N\theta$. In particular, Fig. 7a shows the variation of $\eta$ with respect to *N* (and $\alpha$) for different values of $\Omega = D/\lambda$. Our first observation is that for low values of $D/\lambda$ there is total transmission (i.e., $\eta = 1.0$). This is due to the fact that the microstructure responds mechanically as a homogeneous material when $\lambda \gg D$. $\eta$ shows a significant reduction for values of $D/\lambda > 0.1$. For small number of layers ($N < 3$), the fraction of transmitted energy decreases almost linearly as a function of *N*. For most values of $D/\lambda$, $\eta$ increases for $N > 3$ until it stabilizes for $N > 10$ ($d < D/10$) to a constant value which is independent of the number of layers. This constant value of $\eta$ is plotted as a function of $D/\lambda$ (inset, Fig. 4a) where it is shown that $\eta$ is minimized for a certain range of $D/\lambda$ and increases asymptotically back to $\eta = 1.0$ for larger values of $D/\lambda$. This means that the microstructure ceases to be effective in filtering the impact energy for $D/\lambda > 2$.

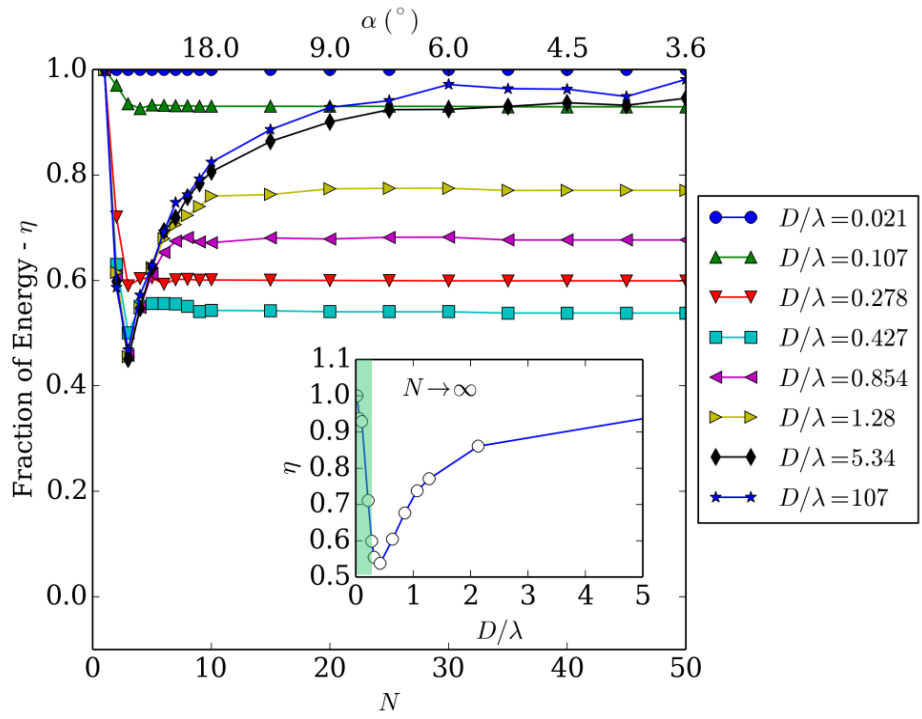

(a)

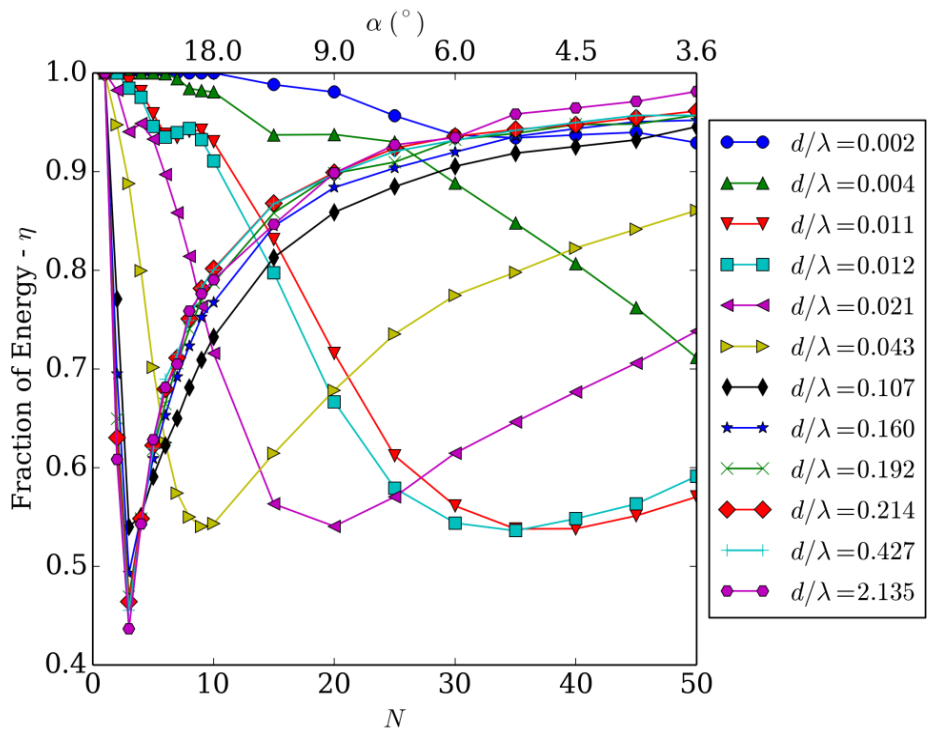

(b)

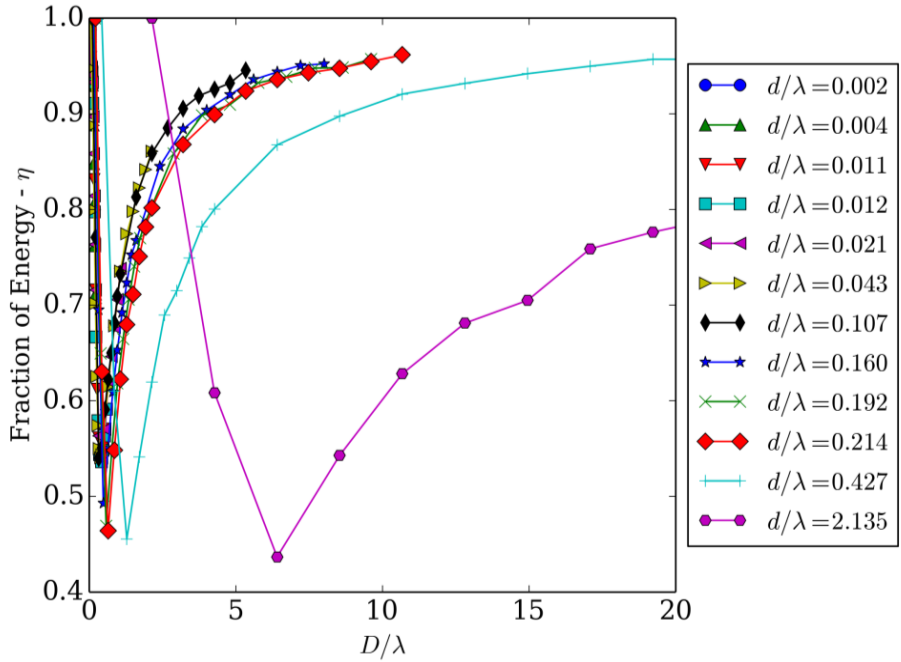

(c)

**Figure 7:** a) Fraction of transmitted energy, $\eta$ vs. $N$ for various values of $D/\lambda$. The inset plot shows $\eta$ for $N \to \infty$. For comparison purposes the range of the stomatopod microstructure is indicated by the shaded area. b) $\eta$ vs. $N$ for various values of $d/\lambda$. c) $\eta$ vs. $D/\lambda$ for various values of $d/\lambda$.

Figure 7b shows the variation of $\eta$ with respect to $N$ (and $\alpha$) for different values of $d/\lambda$. In this case, $D$ increases with $N$ and since each specific plot corresponds to a constant value of the layer thickness $d$, a constant value of $N$ implies a different value for the pitch distance, $D$. A direct comparison between these plots and those in Fig. 7a suggest that performing variations in the parameter $D/\lambda$ seems more effective in the control of the fraction of transmitted energy $\eta$. This is especially relevant if we want to extend this concept to manmade fiber-reinforced composites, where the prepreg thickness (typically predetermined by the manufacturer) limits the choice of $d/\lambda$. Thus, given a particular thickness $d$, one can choose the pitch angle $\alpha$ that is required to achieve a targeted $\eta$. This analysis in terms of fraction of transmitted energy indicates that there is a stronger dependence of the transmitted energy on the dimensionless pitch distance $D/\lambda$. Finally, Figure 7c shows the variation of $\eta$ with $D/\lambda$ for different values of $d/\lambda$. It should be noted that Fig. 7c provides the same information plotted in Figs. 7a and b. Variations of $D/\lambda$ also indicated variations in $N$ and $\alpha$ for constant values of $d/\lambda$. However, this figure shows that when $d/\lambda < 0.2$, most of the $\eta$ vs. $D/\lambda$ curves are similar and follow a similar 'master' curve. Deviation from this curve takes place when $d/\lambda > 0.2$, where it is possible that the wave begins to interact with the characteristic microstructural length $\ell$.

Given the uncertainty on these properties, we employ a range of $v_{12}$ and $v_{23}$ from 0.25 to 0.3, and computed $\eta(D,N)$. To see the differences we computed the relative error between the two functions, i.e.

$$\text{error} = \frac{\|\eta(D,N;\upsilon=0.25) - \eta(D,N;\upsilon=0.3)\|_2}{\|\eta(D,N;\upsilon=0.25)\|_2}$$

what gives a relative error of 0.39 %. For the minimum of the curve (0.7) this represents a 0.12 % of the total energy, showing that the transmitted energy is not that sensitive to changes in the Poisson ratio. In the case of the shear modulus, we varied it from 0.35 GPa to 0.7 GPa (keeping the same order of magnitude) and computed $\eta(D,N)$ in each case. One more time, the relative error was computed as

$$\text{error} = \frac{\|\eta(D,N;G=0.35\text{ GPa}) - \eta(D,N;G=0.7\text{ GPa})\|_2}{\|\eta(D,N;G=0.7\text{ GPa})\|_2}$$

what gives a relative error of 8.75 %. And, for the minimum of the curve (0.7) this represents 2.63 % of the total energy.

## 4. Conclusions

Using the model of a periodic Bouligand-like structure identified by Weaver et al. [7] in the Stomatopod's dactyl club, we have conducted a study intended to elucidate some of the physical mechanisms responsible for the amazing capabilities of the stomatopods dactyl clubs to sustain high intensity dynamical impacts assuming that the material remains elastic during every single impact. For that purpose, we combined the propagator matrix formalism to represent the layered helicoidal structure, with Bloch-Floquet periodic boundary conditions to account for the spatial periodicity. Using this approach, we were able to identify a dispersive response, with related bandgaps, for propagation modes related to shear waves. These frequency bandgaps were also shown to correspond with the characteristic frequency bands generated from the stress pulse experienced by the dactyl club during the stomatopod's hunting activities. In order to quantify the potential filtering effect of such periodic microstructure, we have also conducted a parametric analysis to study the variation of the fraction of transmitted energy with the number of layers that conforms a representative material cell. We varied the ratio of pitch distance to incident wavelength, $D/\lambda$, and the ratio between the layer thickness and the incident wavelength, $d/\lambda$. These values undoubtedly govern the number of layers $N$ and pitch angle $a$, which are part of the microstructural design. For values of $D/\lambda$ between 0.005 - 0.23 and for a number of layers sufficiently large ($N >10$), which is presumed to be found within the dactyl club, we found fractions of transmitted energy in the range 0.7 - 1.0. This regime is indicated as shaded area in the inset figure of Fig. 7a. It is thus concluded that in addition to possible inelastic and damage effects, the dactyl club appears to have the ability to sustain high intensity dynamical loads through a shear wave filtering capability introduced by the periodic nature combined with the chirality of its hierarchically arranged microstructure. Lessons from this study could provide relevant design guidelines for the fabrication of biomimetic impact-tolerant fiber-reinforced composite materials. Future work will focus on employing these models to make predictions and optimize composite designs for very specific applications. Certainly, how efficient the composite will be to absorb energy will depend on the impact conditions. As such, optimum pitch angle, $D$ and $d$ can be determined if the frequency spectrum of the impact is fully understood.

## Acknowledgments

The authors wish to acknowledge financial support from the National Science Foundation through the CAREER award CMMI 1254864 and the Air Force Office of Scientific Research (AFOSR-FA9550-12-1-0245).

## References

[1] L. Grunenfelder, N. Suksangpanya, C. Salinas, G. Milliron, N. Yaraghi, S. Herrera, K. Evans-Lutterodt, S. Nutt, P. Zavattieri, and D. Kisailus, "Bio-Inspired Impact Resistant Composites," *Acta Biomaterialia*, vol. 10, no. 9, pp. 3997–4008, 2014.

[2] M. A. Meyers, J. McKittrick, and P.-Y. Chen, "Structural Biological Materials: Critical Mechanics-Materials Connections," *science*, vol. 339, no. 6121, pp. 773–779, 2013.

[3] U. Wegst and M. Ashby, "The mechanical efficiency of natural materials," *Philosophical Magazine*, vol. 84, no. 21, pp. 2167–2186, 2004.

[4] J. Marshall, T. W. Cronin, and S. Kleinlogel, "Stomatopod eye structure and function: A review," *Arthropod Structure & Development*, vol. 36, no. 4, pp. 420–448, 2007.

[5] S. Patek, W. Korff, and R. Caldwell, "Biomechanics: Deadly strike mechanism of a mantis shrimp," *Nature*, vol. 428, no. 6985, pp. 819–820, 2004.

[6] H. H. Thoen, M. J. How, T.-H. Chiou, and J. Marshall, "A Different Form of Color Vision in Mantis Shrimp," *Science*, vol. 343, no. 6169, pp. 411–413, 2014.

[7] J. C. Weaver, G. W. Milliron, A. Miserez, K. Evans-Lutterodt, S. Herrera, I. Gallana, W. J. Mershon, B. Swanson, P. Zavattieri, E. DiMasi, and others, "The stomatopod dactyl club: a formidable damage-tolerant biological hammer," *Science*, vol. 336, no. 6086, pp. 1275–1280, 2012.

[8] W. Brooks, "Report on the Scientific Results of the Exploring Voyage of the HMS Challenger, vol. 45," *Stomatopoda.(Neill and Co., Edinburgh, 1886)*.

[9] J. Currey, A. Nash, and W. Bonfield, "Calcified cuticle in the stomatopod smashing limb," *Journal of Materials Science*, vol. 17, no. 7, pp. 1939–1944, 1982.

[10] S. Patek and R. Caldwell, "Extreme impact and cavitation forces of a biological hammer: strike forces of the peacock mantis shrimp Odontodactylus scyllarus," *Journal of experimental biology*, vol. 208, no. 19, pp. 3655–3664, 2005.

[11] R. L. Caldwell and H. Dingle, "Stomatopods," *Scientific American*, vol. 234, pp. 80–89, 1975.

[12] B. Banerjee, *An Introduction to Metamaterials and Waves in Composites*. CRC Press, 2011.

[13] F. Capolino, *Theory and phenomena of metamaterials*. CRC Press, 2009.

[14] N. Engheta and R. W. Ziolkowski, *Metamaterials: physics and engineering explorations*. John Wiley & Sons, 2006.

[15] P. Wang, F. Casadei, S. Shan, J. C. Weaver, and K. Bertoldi, "Harnessing Buckling to Design Tunable Locally Resonant Acoustic Metamaterials," *Physical Review Letters*, vol. 113, no. 1, p. 014301, 2014.

[16] P. A. Deymier, Ed., *Acoustic metamaterials and phononic crystals*, 1st ed., vol. 173. Springer, 2013, p. 378.

[17] M. I. Hussein, M. J. Leamy, and M. Ruzzene, "Dynamics of phononic materials and structures: Historical origins, recent progress, and future outlook," *Applied Mechanics Reviews*, vol. 66, no. 4, p. 040802, 2014.

[18] L. Brillouin, *Wave propagation in periodic structures: electric filters and crystal lattices*. Courier Dover Publications, 2003.

[19] C. Kittel and P. McEuen, *Introduction to solid state physics*, vol. 8. Wiley New York, 1986.

[20] J. D. Joannopoulos, S. G. Johnson, J. N. Winn, and R. D. Meade, *Photonic crystals: molding the flow of light*. Princeton university press, 2011.

[21] A.-C. Hladky-Hennion and J.-N. Decarpigny, "Analysis of the scattering of a plane acoustic wave by a doubly periodic structure using the finite element method: Application to Alberich anechoic coatings," *J. Acoust. Soc. Am*, vol. 90, pp. 3356–3367, 1991.

[22] M. Ruzzene, F. Scarpa, and F. Soranna, "Wave beaming effects in two-dimensional cellular structures," *Smart Materials and Structures*, vol. 12, pp. 363–372, 2003.

[23] E. A. Zimmermann, B. Gludovatz, E. Schaible, N. K. Dave, W. Yang, M. A. Meyers, and R. O. Ritchie, "Mechanical adaptability of the Bouligand-type structure in natural dermal armour," *Nature communications*, vol. 4, p. 2634, 2013.

[24] L. Cheng, L. Wang, and A. M. Karlsson, "Mechanics-based analysis of selected features of the exoskeletal microstructure of Popillia japonica," *Journal of Materials Research*, vol. 24, no. 11, pp. 3253–3267, 2009.

[25] P.-Y. Chen, A. Y.-M. Lin, J. McKittrick, and M. A. Meyers, "Structure and mechanical properties of crab exoskeletons," *Acta Biomater.*, vol. 4, no. 3, pp. 587–596, 2008.

[26] A. Al-Sawalmih, C. Li, S. Siegel, H. Fabritius, S. Yi, D. Raabe, P. Fratzl, and O. Paris, "Microtexture and chitin/calcite orientation relationship in the mineralized exoskeleton of the American lobster," *Advanced Functional Materials*, vol. 18, no. 20, pp. 3307–3314, 2008.

[27] C. Sachs, H. Fabritius, and D. Raabe, "Influence of microstructure on deformation anisotropy of mineralized cuticle from the lobster Homarus americanus," *Journal of Structural Biology*, vol. 161, no. 2, pp. 120–132, 2008.

[28] C. Sachs, H. Fabritius, and D. Raabe, "Experimental investigation of the elastic–plastic deformation of mineralized lobster cuticle by digital image correlation," *Journal of Structural Biology*, vol. 155, no. 3, pp. 409–425, 2006.

[29] C. Sachs, H. Fabritius, and D. Raabe, "Hardness and elastic properties of dehydrated cuticle from the lobster Homarus americanus obtained by nanoindentation," *Journal of materials research*, vol. 21, no. 08, pp. 1987–1995, 2006.

[30] S. Lhadi, S. Ahzi, Y. Remond, S. Nikolov, and H. Fabritius, "Effects of homogenization technique and introduction of interfaces in a multiscale approach to predict the elastic properties of arthropod cuticle," *Journal of the mechanical behavior of biomedical materials*, vol. 23, pp. 103–106, 2013.

[31] S. Nikolov, H. Fabritius, M. Petrov, M. Friák, L. Lymperakis, C. Sachs, D. Raabe, and J. Neugebauer, "Robustness and optimal use of design principles of arthropod exoskeletons studied by ab initio-based multiscale simulations," *Journal of the mechanical behavior of biomedical materials*, vol. 4, no. 2, pp. 129–145, 2011.

[32] S. Nikolov, M. Petrov, L. Lymperakis, M. Friák, C. Sachs, H.-O. Fabritius, D. Raabe, and J. Neugebauer, "Revealing the Design Principles of High-Performance Biological Composites Using Ab initio and Multiscale Simulations: The Example of Lobster Cuticle," *Advanced Materials*, vol. 22, no. 4, pp. 519–526, 2010.

[33] H. Hepburn and A. Ball, "On the structure and mechanical properties of beetle shells," *Journal of Materials Science*, vol. 8, no. 5, pp. 618–623, 1973.

[34] P. Vukusic and J. R. Sambles, "Photonic structures in biology," *Nature*, vol. 424, no. 6950, pp. 852–855, 2003.

[35] C. Campos-Fernández, D. E. Azofeifa, M. Hernández-Jiménez, A. Ruiz-Ruiz, and W. E. Vargas, "Visible light reflection spectra from cuticle layered materials," *Optical Materials Express*, vol. 1, no. 1, pp. 85–100, 2011.

[36] P. Zhang and A. C. To, "Broadband wave filtering of bioinspired hierarchical phononic crystal," *Applied Physics Letters*, vol. 102, no. 12, pp. 121910–121910, 2013.

[37] B. A. Auld, *Acoustic fields and waves in solids*, vol. 1. Wiley New York, 1973.

[38] V. Buchwald, "Elastic waves in anisotropic media," *Proceedings of the Royal Society of London. Series A. Mathematical and Physical Sciences*, vol. 253, no. 1275, pp. 563–580, 1959.

[39] J. J. M. Carcione, *Wave fields in real media: Wave propagation in anisotropic, anelastic and porous media*, vol. 31. Pergamon, 2001.

[40] S. K. Yang, V. V. Varadan, A. Lakhtakia, and V. K. Varadan, "Reflection and transmission of elastic waves by a structurally chiral arrangement of identical uniaxial layers," *Journal of Physics D: Applied Physics*, vol. 24, no. 9, p. 1601, 1991.

[41]  R. F. Gibson, *Principles of composite material mechanics*. CRC Press, 2011.
[42]  V. Varadan, S. Yang, and V. Varadan, "Rotation of elastic shear waves in laminated, structurally chiral composites," *Journal of sound and vibration*, vol. 159, no. 3, pp. 403–420, 1992.
[43]  L. Cheng, "Mechanical implications of the arthropod exoskeleton microstructures and the mechanical behavior of the bioinspired composites," University of Delaware, 2010.
[44]  L. Cheng, L. Wang, and A. M. Karlsson, "Image analyses of two crustacean exoskeletons and implications of the exoskeletal microstructure on the mechanical behavior," *J. Mater. Res.*, vol. 23, no. 11, pp. 2854–2872, Nov. 2008.
[45]  R. G. Payton, *Elastic wave propagation in transversely isotropic media*, 1st ed., no. 4. Martinus Nijhoff Publishers, 1983, p. 192.
[46]  C. Oldano and S. Ponti, "Acoustic wave propagation in structurally helical media," *Physical Review E*, vol. 63, no. 1, p. 011703, 2000.
[47]  F. Gilbert and G. E. Backus, "Propagator matrices in elastic wave and vibration problems," *Geophysics*, vol. 31, no. 2, pp. 326–332, 1966.
[48]  P. Helnwein, "Some remarks on the compressed matrix representation of symmetric second-order and fourth-order tensors," *Computer methods in applied mechanics and engineering*, vol. 190, no. 22, pp. 2753–2770, 2001.
[49]  N. Haskell, "The dispersion of surface waves on multilayered media," *Bulletin of the Seismological Society of America*, vol. 43, no. 1, pp. 17–34, 1953.
[50]  H. Hochstadt, *Differential equations: a modern approach*. Courier Dover Publications, 1975.
[51]  L. Brillouin, *Wave propagation and group velocity*, vol. 960. Academic Press New York, 1960.